\documentclass[prl,twocolumn,showpacs,preprintnumbers,amsmath,amssymb]{revtex4-2}

\usepackage{graphicx}
\usepackage{dcolumn}
\usepackage{bm}

\DeclareMathOperator{\erfc}{erfc}
\DeclareMathOperator{\sgn}{sgn}

\begin{document}

\newcommand{\spinup}{\protect{$ \left|\uparrow \right\rangle$}}
\newcommand{\spindown}{\protect{$ \left|\downarrow \right\rangle$}}
\newcommand{\upeq}{\protect{\left | \uparrow \right\rangle}}
\newcommand{\downeq}{\protect{\left | \downarrow \right\rangle}}

\title{Spectroscopic Characterization of the Quantum Linear-Zigzag Transition in Trapped Ions}

\author{J. Zhang$^{1,2}$, B.~T. Chow$^1$, S. Ejtemaee$^1$, and P.~C. Haljan$^1$}
\affiliation{$^1$Department of Physics, Simon Fraser University,
Burnaby, BC, V5A 1S6, Canada,}
\affiliation{$^2$Department of Physics, National University of Defense Technology, Hunan, China.}

\date{\today}

\begin{abstract}
While engineered quantum systems are a general route to the manipulation of multipartite quantum states, access in a physical system to a continuous quantum phase transition under sufficient control offers the possibility of an intrinsic source of entangled states. To this end we realize the quantum version of the linear-zigzag structural transition for arrays of up to five ground state-cooled ions held in a linear Paul trap and we demonstrate several of the control requirements towards entangled-state interferometry near the critical point. Using in-situ spectroscopy we probe the energy level structure and occupation of the soft mode associated with the structural transition, and show a stable critical point and majority ground state occupation crossing the transition. We resolve biases arising from trap electrode asymmetries that change the nature of the transition, show that they can be suppressed by varying the ion number, and demonstrate control of the transition bias using optical dipole forces.
\end{abstract}


\maketitle

By virtue of their Coulomb interactions, laser-cooled trapped arrays of ions intrinsically present a strongly interacting condensed matter system that forms a variety of Wigner ion-crystal configurations~\cite{Diedrich1987a, Wineland1987a, Birkl1992a, Mitchell1998a, CPTvol1} and at the same time is dilute enough to be optically manipulated down to single atom resolution. The ion-crystal configurations are separated by a hierarchy of structural phase transitions driven by either the confinement geometry or ion density~\cite{Raizen1992a, Birkl1992a, Mitchell1998a, Schiffer1993a, Walther1995a, Dubin1999a}. In a linear radio-frequency Paul trap the first such transition is the 1D linear to 2D zigzag transition~\cite{Raizen1992a, Schiffer1993a}, which for small arrays of ions is the mesoscopic analog of a continuous phase transition~\cite{Piacente2004a, Fishman2008a}. Prior experiments with the linear-zigzag (LZ) transition have focused on its classical behaviour including both equilibrium properties~\cite{Raizen1992a, Enzer2000a, Dubin1999a} and dynamics~\cite{Liang2011a, Ejtemaee2013a, Pyka2013a, Ulm2013a, Kiethe2017a}. Here, using ground-state cooling we investigate the transition in the quantum regime, and we assess the feasibility of the system for double-well interferometry~\cite{Retzker2008a} and the sensing of ambient electric field noise in the ion-trap environment. As an important step towards interferometry, we investigate the near-adiabatic crossing of the LZ transition in the ground state, which in the ideal scenario prepares a Schrodinger cat superposition of the symmetry broken zigzag structures.

The predominant paradigm underlying quantum control of trapped arrays of ions, including in quantum computing and quantum simulations, uses the set of vibrational normal modes of the ion crystal in the linearized small oscillation limit~\cite{Blatt2008a}. Near the LZ critical point this work explores the opposite limit where the nonlinearity in the interactions dominates the effective potential of the relevant zigzag vibrational mode. As a further contrast, dynamical Schrodinger cat states involving entangled states of spin and coherent motion~\cite{Monroe1996a} and generalized Greenberger–Horne–Zeilinger (GHZ) states of spin~\cite{Leibfried2005a, Monz2011a} have both been previously prepared in ion traps. Here, the limit of quantum state preparation in an ion trap can be explored down to dc excitation frequencies due to the softening of the zigzag mode at the LZ transition, and coherent state manipulation on the zigzag side of the transition would provide a testbed to probe sources of static and fluctuating bias affecting the double-well zigzag potential~\cite{Retzker2008a}. Unlike prior quantum dynamics studies near the critical point for a three-ion rotor mode~\cite{Noguchi2014a}, the LZ transition is a system that is readily extensible to a varied and larger number of ions, and in-situ measurement of near ground-state energies is shown here to be feasible close to the critical point.

Our experiments are simultaneously motivated by the fact that the LZ transition at ultracold temperatures is marked by a sharp spectral signature in the zigzag mode with strong dependence on the trap potential parameters. This allows for sensitive in-situ and broadband electric field noise measurement, which is of direct consequence to the performance of trapped ion quantum computers~\cite{Blatt2008a, Bruzewicz2019a}, and offers advantages over the standard sensing technique based on the center-of-mass mode of a single trapped ion~\cite{Turchette2000a, Brownnutt2015a}. First, the measurement of slow drifts in the strength of the ion-trap potential can be achieved with more than an order-of-magnitude improvement in single-shot sensitivity without the need for high-order motional Fock-state superpositions~\cite{Mccormick2019a}. Second, as a resonant absorptive sensor the zigzag mode near the LZ transition offers wide frequency tunability from dc to 1~MHz for only minimal adjustment of the trap voltages ($\sim$1 V), which is advantageous in the characterization of the spectral dependence of noise to identify its sources~\cite{Brownnutt2015a}.

\begin{figure}[t]
\centering
\includegraphics[width=1.0\linewidth,clip]{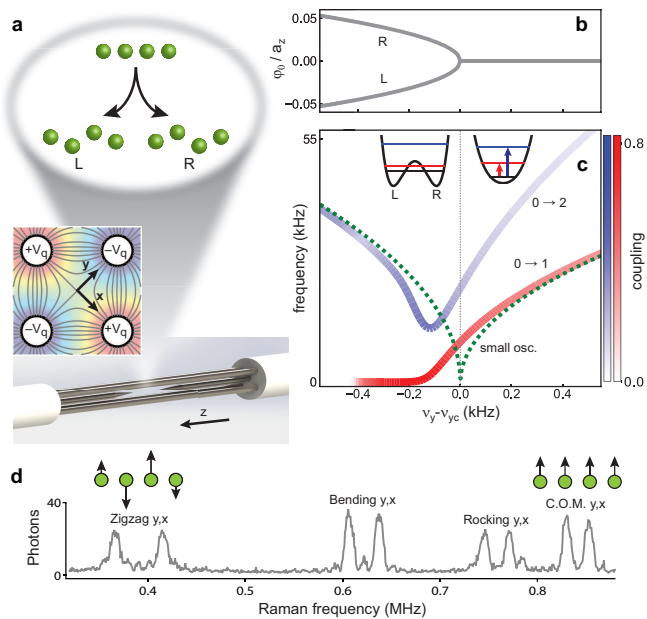}
\caption{\textbf{Linear-zigzag transition.} (a) Schematic realization of the 1D linear to 2D zigzag structural transition for a four-ion crystal confined in a linear rf Paul trap, consisting of four rod electrodes and two endcap needles as shown. Color inset shows an end view of the trap superposed with the transverse quadrupole electric field generated from rod voltages $\pm V_q$ and used to weaken the transverse confinement along the $y$-axis to induce the transition. (b) Classical energy-minimum values of the zigzag order parameter, proportional to the transverse zigzag displacement from the $z$-axis and scaled to the axial Coulomb length scale $a_z$, as a function of control parameter $\nu_y-\nu_{yc}$. $\nu_y$ is the transverse-$y$ secular trap frequency; $\nu_{yc}=740$ kHz and $a_z=5.4\,\mu$m for the four-ion simulation shown. (c) Simulated quantum energy-level spectrum for the first two excited states of the transverse-$y$ zigzag mode of a four-ion crystal, measured relative to the ground state. A small linear bias $C_1=10^{-14}$ is assumed. Color shading of the levels indicates the strength of Raman sideband coupling to the ground state for a central ion (see sidebar scale). Also shown is the classical small-oscillation prediction and the form of the zigzag potential on either side of the transition. (d) Experimental Raman sideband spectrum of all transverse vibrational modes for a four-ion crystal. A Raman probe of the center-of-mass mode provides \textit{in-situ} assessment of the control parameter, while a probe of the zigzag mode provides access to order-parameter properties including energy-level spectrum, level occupancy and coherence.} \label{fig:setup}
\end{figure}

Assuming that one transverse axis of the ion trap is tightly confining, we can describe the 2D dimensionless potential for $N$ ions in a linear radio-frequency (rf) Paul trap as
\begin{equation}\label{eqn:2Dpotential}
V(y, z) = \sum_{i=1}^N{\frac{1}{2}\left(\alpha y_i^2 + z_i^2\right)}
+\sum_{i<j}^n{ \frac{1}{ |\vec{r}_i - \vec{r}_j| }  }
+ V_{pert}(y, z)
\end{equation}

\noindent which includes the harmonic rf pseudopotential, Coulomb interactions and perturbative imperfections $V_{pert}$ in the trap potential. The aspect ratio of the harmonic trap is characterized by the quantity {\protect$\alpha = (\omega_y/\omega_z)^2$} in terms of the secular trap frequencies {\protect$\nu_i=\omega_i/2\pi$}. The dimensionless potential is expressed in terms of the Coulomb length $a_z=\left(\frac{q^2}{4\pi\epsilon_0m\omega_z^2}\right)^{1/3}$, which sets the axial ion spacing, and the corresponding Coulomb energy $\frac{q^2}{4\pi\epsilon_0a_z}$. The perturbation potential captures the effect of deformations of the electrode geometry away from an ideal linear trap, with select terms in a polynomial expansion leading to symmetry breaking of the LZ transition.

The LZ transition for an ion crystal with fixed number of ions is controlled by the trap aspect ratio $\alpha$. This can be modified by applying a dc quadrupole potential in the transverse trapping plane to weaken the transverse confinement along one of the principal axes, here assumed to be the $y$-axis (Fig.~\ref{fig:setup}a). For strong transverse confinement the ions form a linear string along the axial $z$-direction of the linear trap. At a critical point $\alpha_c$ which depends on the number of ions~\cite{Enzer2000a}, the ions undergo a structural phase transition to a 2D zigzag configuration (Fig.~\ref{fig:setup}a). The dynamics of the linear ion string restricted to 2D can be described in terms of its $2N$ collective vibrational modes. The transverse-$y$ zigzag mode represents a ``soft mode"~\cite{Cowley2012a} that classically goes to zero frequency at the LZ critical point according to $\nu_{zz}=\nu_z\sqrt{\alpha-\alpha_c}$~\cite{Fishman2008a}. This creates a dynamical instability that drives the transition~\cite{Enzer2000a}, and below the critical point the crystal's equilibrium structure takes on a frozen-in version of the zigzag mode~\cite{Fishman2008a, Cowley2012a}. Near the transition the zigzag mode dominates the ion crystal dynamics for slow quenches. A coupled-mode analysis and adiabatic elimination of the other modes leads to an effective field theory for the zigzag mode~\cite{Retzker2008a, Chow2022a} (see Methods). The associated dimensionless potential as a function of the zigzag order parameter $\varphi$ (the normal mode coordinate) up to fourth order is

\begin{equation}
U(\varphi)=C_1\varphi + \frac{1}{2}C_2\varphi^2 + \frac{1}{3}C_3\varphi^3 + \frac{1}{4}C_4\varphi^4
\label{eqn:Ueff}
\end{equation}

\noindent where the quadratic term is $C_2 = \alpha - \alpha_c$, and other coefficients are constant. This derivation ignores spatial variation of the order parameter and propagation effects along the ion string~\cite{Shimshoni2011a, Chiara2010a}, which are not relevant for the small ion strings and slow quenches considered here. Asymmetries in the non-ideal trap from $V_{pert}$ give rise to the linear and cubic bias terms. If $C_1$ and $C_3$ are zero, the mesoscopic equivalent of a second-order phase transition is realized for small numbers of ions. The mean order parameter $\langle\varphi\rangle$, obtained from minimization of $U$, is zero on the linear side and continuously acquires a non-zero value on the zigzag side of critical point (Fig.~~\ref{fig:setup}b). On the zigzag side of the transition there are two symmetry-broken, and in the ideal case energy-degenerate, states corresponding to the minima of the quartic double-well potential that forms across the transition. We define the two equilibrium configurations as ``left" (\textit{L}) and ``right" (\textit{R}). In the zero temperature limit, it is relevant to consider the quantum energy levels for the effective potential across the transition, as shown in Fig.~\ref{fig:setup}d for the example of four ions. Near the critical point the frequency splitting between the ground state, $|0\rangle$, and first excited state, $|1\rangle$ deviates from the equivalent classical small oscillation frequency. The level splitting remains finite at the critical point before approaching zero in the zigzag phase as tunnel coupling between the two sides of the double well, associated with the states $|L\rangle\equiv\left(|0\rangle - |1\rangle\right)/\sqrt{2}$ and $|R\rangle\equiv\left(|0\rangle + |1\rangle\right)/\sqrt{2}$, is suppressed by the intervening barrier. At the point of optimum tunneling the two lowest levels are just captured below the double-well barrier. For $\nu_y=0.75$ MHz a tunnel splitting of 3 kHz is expected at this point, and the order parameter's magnitude is $|\langle\varphi\rangle|=0.03-0.02$ or 100--50 nm for 3--5 ions.

Small nonlinearities in the trap potential from the electrode configuration or from other ambient sources will introduce biases changing the nature of the phase transition. A small cubic term will change the transition from second to weak first order~\cite{Plischke2006}. A linear bias will act to smooth away the discontinuity in the transition. From the perspective of double-well interferometry and entangled state preparation, we seek to realize a sufficiently symmetric double well such that quantum tunneling is not suppressed near the critical point. To estimate the relevant level of bias, we consider the optimum tunneling point for 3--5 ions where a bias in the ground-state wavefunction limited to the range $0.5 < |\langle R|0\rangle|^2 < 0.75$ requires a potential bias of $|C_1|\lesssim5\times10^{-7}$ or $|C_3|\lesssim5\times10^{-3}$. At the same time, stability of the potential is required together with sufficiently low noise to retain coherence.

\section{\label{sec:results} Results\protect\\}

\noindent\textbf{Experimental system.} Our investigations of the LZ transition use $^{171}$Yb$^+$ ions held in a stabilized linear radio-frequency (rf) Paul trap~\cite{EjtemaeePhD2015, Ejtemaee2013a} (Fig.~\ref{fig:setup}a), which has typical secular frequencies $\left\{\nu_{x0}, \nu_{y0}, \nu_{z0}\right\}$ of $\left\{864, 844, 303\right\}$ kHz for five ions starting on the linear side of the transition (See Methods for further experimental details). Near-ground-state cooling of the linear-string configuration is achieved through 3D Sisyphus cooling of all $3N$ vibrational modes to the few-phonon level~\cite{Ejtemaee2013a} followed by simultaneous resolved sideband cooling of the transverse-$y$ and axial modes except the center-of-mass (COM) ones. We estimate a ground-state occupation $\gtrsim0.9$ for the sideband cooled modes including the transverse-$y$ zigzag mode of interest~\cite{Ejtemaee2017a,EjtemaeePhD2015}. Following the cooling process, the approach to and crossing of the LZ transition is controlled by a ramp of a transverse dc quadrupole potential applied through the trap rods (Fig.~\ref{fig:setup}b)~\cite{Ejtemaee2013a,EjtemaeePhD2015} with minimal effect on the axial confinement ($|\Delta \omega_z|/\omega_z < 0.03\%$). The secular frequency along the $y$-axis weakens while the orthogonal $x$-axis simultaneously strengthens such that the LZ transition is effectively confined to the 2D $y-z$ plane. While the ramp implementation is expected to be adiabatic for endpoints near the critical point, we do not optimize the ramp for endpoints deeper in the zigzag phase. At a given final ramp value of the dc quadrupole voltage, the vibrational modes of the ion crystal are probed by driving stimulated two-photon Raman sideband transitions between the internal hyperfine states $\protect{^2\!S_{1/2}|0,0\rangle}\equiv\downeq$ and $\protect{^2\!S_{1/2}|1,0\rangle}\equiv\upeq$ of the ions, separated by $\nu_0\approx12.6$ GHz~\cite{Ejtemaee2017a,EjtemaeePhD2015}. Subsequent readout of the transition is obtained by state-selective fluorescence of the internal state of the ions~\cite{Ejtemaee2010a}. For individual laser addressing of the $i^{th}$ ion, the Raman coupling drives the transition $|\downarrow\rangle^{\bigotimes N}|n_k\rangle \rightarrow |\downarrow...\uparrow_i...\downarrow\rangle|n_k^\prime\rangle$ at resonance $\nu_0+(n_k-n_k^\prime)\nu_k$, involving the vibrational states $|n_k\rangle$ and $|n_k^\prime\rangle$ of the $k^{th}$ mode of the ion crystal.  In practice we use a technically simpler global illumination of the ions, which gives rise to a simultaneous Raman coupling of all the ions to the mode of interest (see Methods). We measure both the carrier ($n_k^\prime=n_k$) and upper sideband resonances (for example the first sideband $n_k^\prime=n_k+1$) and extract the mode frequency from the difference. The first sidebands of the COM modes are used to measure the secular frequencies of the trap -- the control parameter for the transition -- while the sidebands for the transverse zigzag mode give access to the spectral properties of the order-parameter dynamics.

\begin{figure*}[t]
\centering
\includegraphics[width=1.0\linewidth]{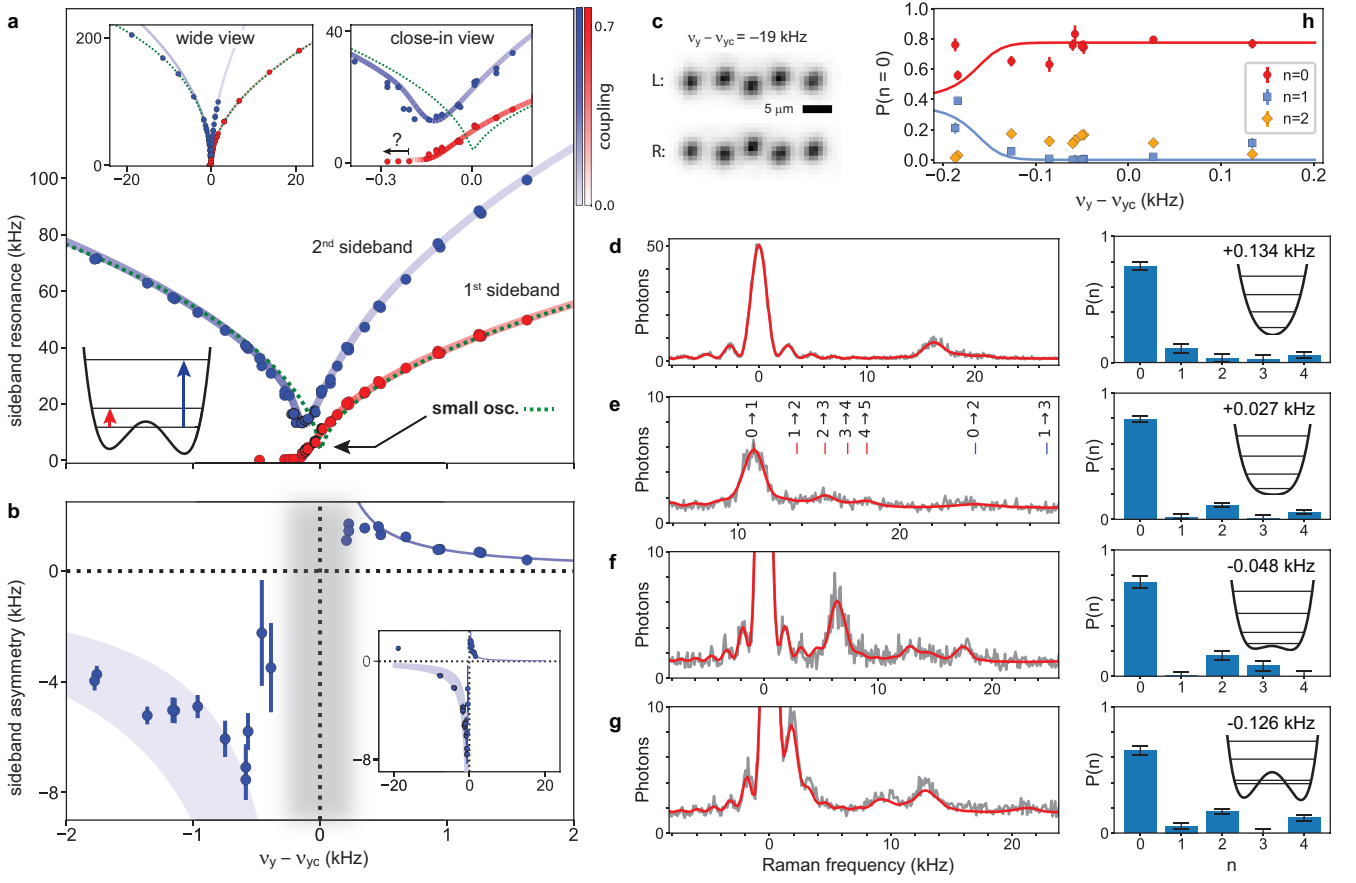}
\caption{\textbf{Raman sideband spectroscopy of the zigzag mode across the linear-zigzag transition for five ions.} (a) Frequency of the first and second upper Raman sidebands for the transverse-$y$ zigzag mode as a function of the secular trap frequency $\nu_y$ relative to the critical value $\nu_{yc}\approx760$ kHz. The $303$-kHz axial secular frequency is nearly constant ($0.03\%$ variation). Insets show the full range of data acquisition and a close-in view near the critical point. Solid lines are quantum energy-level differences of the $n=1$ and $n=2$ number states with respect to the $n=0$ ground state for a quartic potential (Eqn.~\ref{eqn:Ueff} ) with linear bias of $|C_1|=3.3\times10^{-7}$. Line shading corresponds to Raman coupling strength (see side-bar scale). Green dotted line shows the corresponding classical small-oscillation prediction. (b) Raman lineshape asymmetry for the second upper sideband. Inset shows the full range of data acquisition. Blue lines indicate the expected scaling of the asymmetry due to the anharmonicity of the quartic potential. The width of the line on the zigzag side shows a factor-of-two range in scaling prefactor. Vertical gray shaded region indicates where sideband peaks are (partially) resolved as shown in d--g. (c) Sample images of the two symmetry-broken equilibrium structures of the five-ion crystal far from the critical point. (d--g) Sample Raman sideband lineshapes near the critical point and fits (solid red lines) to extract $C_1$ and Fock-state population distribution $P(n)$ of the zigzag mode. Motional distributions from the fits are shown in adjacent panels along with the expected shape of the zigzag potential. In (e) line centers for sidebands from different initial $n$--levels are indicated. (h) Summary of measured $P(n)$ across the transition for $n\leq2$. Lines are a quantum simulation for an initial ground state to indicate onset of non-adiabaticity. Simulation probabilities are weighted to match the average of $P(0)$ data on the linear side of the transition. The lowest two energy levels in the double-well model lie below the barrier near $\nu_y-\nu_{yc}=-0.100$ kHz.}
\label{fig:transitionSpec}
\end{figure*}

\vspace{0.5em}
\noindent\textbf{Transition spectroscopy.} We first consider the properties of the LZ transition in the vicinity of the critical point for a crystal of five ions. For this study both the first and second upper sidebands of the zigzag mode are measured as a function of secular frequency $\nu_y$. A new feature from the typical behaviour for linear ion strings is the significant anharmonicity of the zigzag mode near the critical point due to the nonlinearities in the effective potential. This gives rise to non-uniform energy level spacings and asymmetric lineshapes. We avoid significant induced distortion of the lineshapes by limiting the drive strength of the Raman transition. Nevertheless, if the zigzag mode is in an initial distribution of Fock states, for example a thermal distribution, this will lead to an asymmetric lineshape. In cases of significant asymmetry, we fit the sideband resonances with an exponentially modified Gaussian to extract a resonance, width and lineshape asymmetry factor (see Methods). Assuming an initial motional distribution with a large ground state occupation, we can associate the first sideband resonance with the $\protect{0-1}$ energy level spacing and the second sideband with the $\protect{0-2}$ energy level spacing in the potential for the zigzag mode.

Figure~\ref{fig:transitionSpec}a shows both first and second sideband resonances of the zigzag mode in a 4--kHz range around the critical point at $\nu_{yc}=759.94(2)$ kHz. The data including the region close to the transition is fit to the quantum energy level theory for a biased quartic potential (see Methods) and provides the value of the critical point used to set the origin of the plot. Expressed as a trap asymmetry, the critical value, $\alpha_c = 6.3007(16)$, is shifted upward from the expected pseudopotential value of 6.2374 by $\delta\alpha_c/\alpha_c=0.0100(2)$. The fractional shift is in agreement with an estimate of 0.01050(3) due to the higher order effect of the rf micromotion on the vibrational modes of the ions in the trap (see Methods and~\cite{Landa2012a}). Quoted uncertainties include both statistical error and error associated with calibration of the trap potential. The quadrupole voltage adjustment for points near the transition is equivalent to a variation of 50 ppm of the transverse secular frequency, providing an indirect characterization of the stability of the trap potential over several hours.

The full dataset taken over a wider range of control parameter is shown in the left inset of Fig.~\ref{fig:transitionSpec}a. The analytical prediction $\nu_{zz} = \sqrt{\nu_y^2 - \nu_{yc}^2}$ for the zigzag frequency on the linear side of the transition matches the data well at points away from the critical point. Far from the transition on the zigzag side, the measured sideband frequencies deviate from the energy level spectrum for the perturbative zigzag potential near the critical point -- as expected -- but good agreement is recovered using a model from classical small oscillation analysis for an ion crystal in the pseudopotential. Near the critical point we have a distinct deviation from the classical small oscillation prediction. Only very close to the transition is this clear -- within $\Delta\nu_y\sim\pm0.3$ kHz. The dependence measured for five ions close to the transition approaches that of an unbiased quantum LZ transition with the first sideband approaching zero frequency (compare Fig.~\ref{fig:setup}c). Bias in the double well, which lifts the degeneracy between the lowest energy levels of the zigzag mode, can be identified by a non-zero minimum frequency of the first sideband. However, on the zigzag side of the critical point it becomes difficult to separate the low-frequency first sideband from the carrier; as such, the lowest frequency value that can be reliably measured provides an upper bound on the bias. Suppression of the line strength of the first sideband resonance near the critical point additionally constrains the bias in the double well, and a combined analysis (discussed in detail below) is used to obtain the fit to quantum theory with $C_1 = 3.3\times10^{-7}$ shown in Fig.~\ref{fig:transitionSpec}a.

In addition to the resonance center, the asymmetry in the lineshape of the zigzag sidebands provides insight into the zigzag potential and the motional distribution of the order parameter. The asymmetry of the fit lineshape is shown in Fig.~\ref{fig:transitionSpec}b, where its sign switches across the critical point due to the change in the anharmonicity from the single to double well. The anharmonic energy-level shift in the perturbative limit should scale as $1/\nu_{zz}^2 \propto 1/\Delta\nu_y$~\cite{Muller-Kirsten2012}. The scaling law shows good agreement on the linear side away from the critical point but only a rough match to the behavior on the zigzag side. The 6--12 times larger asymmetry scaling factor inferred on the zigzag side is partly attributed to a factor-of-2 larger anharmonicity effect, and the remainder we expect is due to a larger motional excitation out of the ground state. The onset of increased heating of the zigzag mode across the LZ transition arises from non-adiabatic transitions due to the ramp, higher electric field noise density at low frequencies and enhanced sensitivity to electric field noise as the zigzag structure forms~\cite{Chow2022a}.

In the region closest to the critical point, the anharmonicity of the potential is sufficiently large to allow the zigzag sidebands starting from the lowest vibrational levels to be resolved (Figs.~\ref{fig:transitionSpec}d-g). We take advantage of this to infer the motional population distribution, $P(n)$, in the $n$-number states of the zigzag mode from the measured sideband spectrum of transitions $n \rightarrow n + m$. The fit model of the line strengths and centers is based on the quantum double-well theory including bias, and the carrier and sidebands are fit to an incoherent sum of lineshapes which include three separate phase decoherence parameters for the carrier, first and second sidebands. It is not possible to constrain independent values for the linear and cubic bias from the resolved spectra; for simplicity, we assume a null cubic bias ($C_3=0$) and take the average of $|C_1|$ fit values across the range of data in Fig.~\ref{fig:transitionSpec} to obtain a best-fit value of $|C_1|=3.3(3)\times10^{-7}$. With this value of $C_1$ we obtain the motional population distributions shown in Fig.~\ref{fig:transitionSpec}d-h. The motional populations reveal non-thermal distributions with preferential occupation of even $n$ levels (for example, Fig.~\ref{fig:transitionSpec}e), which is indicative of non-adiabatic transitions due to the ramp or parametric heating of the zigzag mode on the linear side. From the aggregate measurements of motional populations (Fig.~\ref{fig:transitionSpec}h) the ground state is found to remain majority populated ($P_0\gtrsim0.6$) down to $\Delta\nu_{y}=-0.2$ kHz, including the location of the optimum tunneling point near $-0.100$ kHz where the lowest two energy levels in the double-well model first lie below the barrier. From numerical modelling of the quantum mechanical dynamics of the ramp, we expect that it remains adiabatic down to $\Delta\nu_{y}=-0.17$ kHz, which is consistent with the observed behavior of $P(n=1)$ as shown in Fig.~\ref{fig:transitionSpec}h.

The linewidths of the resolved sidebands give an upper bound on the coherence time for the zigzag order parameter. We find that the phase coherence time from the fits for the first sidebands is 0.3--0.5 ms and 0.3 ms for the second sidebands, both much shorter than the value of 3--5 ms for the motion-sensitive carrier. The value of 0.3--0.5 ms is roughly consistent with a preliminary assessment of the Ramsey coherence time of the first upper sideband near to the critical point. We observe that the coherence time decreases as the LZ transition is approached from the linear side. In part this is due to increased sensitivity of the zigzag mode to fluctuations in the trap potential, scaling as $1/\nu_{zz}^2$. A more detailed assessment of decoherence near the critical point is left to future work.

\begin{figure}[t]
\centering
\includegraphics[width=1.0\linewidth]{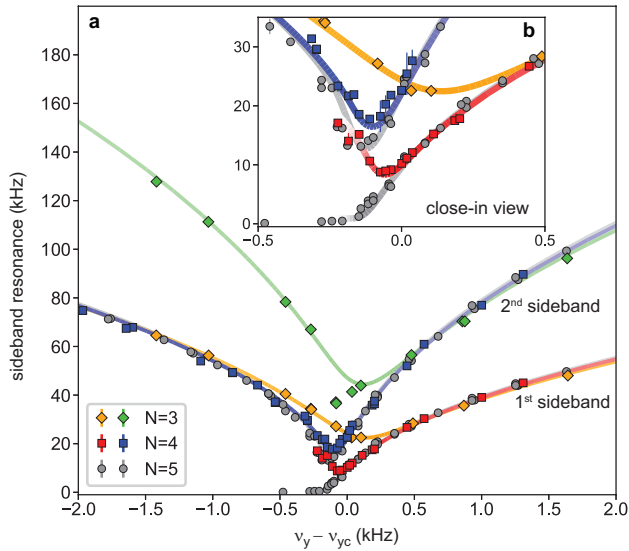}
\caption{\textbf{Comparison of the linear-zigzag transition for 3--5 ions.} First and second upper sideband frequencies for the transverse-$y$ zigzag mode are shown as a function of the transverse secular trap frequency $\nu_y$ referenced to the measured critical value $\nu_{yc}$. Axial confinement is chosen such that the critical values $\nu_{yc}=\{717, 744, 760\}$ kHz for 3--5 ions are close (within 6\%) so that the linear-zigzag (LZ) transitions can be directly compared. Fits (solid lines) to the quantum energy level structure for a quartic potential are used to extract $\nu_{yc}$ and a bias $|C_1|$ of $3.62(11)\times10^{-5}$ for three ions and $4.52(17)\times10^{-6}$ for four ions. For five ions, theory is shown for $|C_1|=3.3(3)\times10^{-7}$, obtained separately from lineshape fits (see text). Inset shows a close-in view of the critical point region. Theory lines are shaded according to Raman coupling for the central ion, similar to Fig.~\ref{fig:transitionSpec}a. In both four- and five-ion cases the $0\rightarrow1$ first sideband coupling vanishes below the LZ transition due to the onset of the symmetry breaking effect on the wavefunctions. This is precluded for three ions since the symmetry is already fully broken by the strong bias.}
\label{fig:NionDependence}
\end{figure}

\vspace{0.5em}
\noindent\textbf{Ion-number dependence.} Extending the results for five ions, we have compared the transition spectroscopy for ion numbers ranging from three to five (Fig.~\ref{fig:NionDependence}). The axial confinement, and by extension the axial ion density, is adjusted to make the critical transverse secular frequency approximately the same in all cases, within a 6\% range. The critical trap asymmetries $\alpha_c$ are found to be \{2.4251(4), 4.1967(6), 6.3007(15)\} for 3--5 ions, corresponding to a nearly $N$-independent fractional shift of \{0.0103(2), 0.01012(14), 0.0100(2)\} from the pseudopotential prediction. This is in accord with the predicted micromotion-induced shift, which in lowest order depends only on the Mathieu parameters for the ion trap~\cite{Chow2022a} (see Methods). The main feature in the transition spectroscopy of Fig.~\ref{fig:NionDependence} is the strong reduction in bias effect going from three to five ions. This is manifest by the qualitative sharping of the LZ transition and the reduction in the gap between the $n=0$ and $n=1$ energy levels near the critical point. Assuming a dominant linear bias -- to be justified below -- we perform fits to the quantum theory for the three- and four-ion cases, and extract a bias coefficient $|C_1|$ of $3.62(11)\times10^{-5}$ and $4.52(17)\times10^{-6}$ respectively. The case of four ions exhibits a complicated spectral structure due to the particular value of the bias; from the best-fit double-well model, we find that the line strength of the $0\rightarrow1$ sideband vanishes as it approaches degeneracy with the $0\rightarrow2$ sideband, corresponding to a level crossing between the $n=1$ and $n=2$ states. For three ions the bias is strong enough that the lowest energy states are well localized in a single well across the LZ transition, and thus neither a suppression of the $0\rightarrow1$ sideband from an onset of symmetry breaking nor a level crossing occur. Finally, we note that we have also confirmed the general trend in bias for $3-5$ ions from measurements of the asymmetry in statistical outcomes of the two zigzag configurations after crossing the LZ transition~\cite{Chow2022a}.

\vspace{0.5em}
\noindent\textbf{Bias spectroscopy.} The transition spectroscopy of Figs.~\ref{fig:transitionSpec} and \ref{fig:NionDependence} is insufficient to distinguish unambiguously between a linear and cubic bias in the zigzag potential. To differentiate the two forms of bias, we introduce a new technique to measure the splitting in the frequency of the zigzag mode between the \textit{L} and \textit{R} zigzag configurations. The splitting dependence on $C_1$ and $C_3$, considering a classical perturbative model derived from Eqn.~\ref{eqn:Ueff} on the zigzag side of the LZ transition and for $|\nu_L-\nu_R|\ll\nu_{L,R}$, is

\begin{equation}\label{eqn:approx_bias_fit_function_vs_freq}
\frac{\nu_{L} - \nu_{R}}{\nu_z} \approx -3C_1\sqrt{2C_4}\frac{\nu_z^2}{\nu_R^2} - \frac{C_3}{\sqrt{2C_4}}, \quad \alpha<\alpha_c
\end{equation}

\noindent where the axial secular frequency $\nu_z$ is used to set the scale. The linear bias is characterized by a rapid $\nu_R^{-2}$ decay away from the transition since it cannot affect the local curvature in the minima of the double well once the wells are deep enough.

\begin{figure}[t!]
\centering
\includegraphics[width=0.96\linewidth]{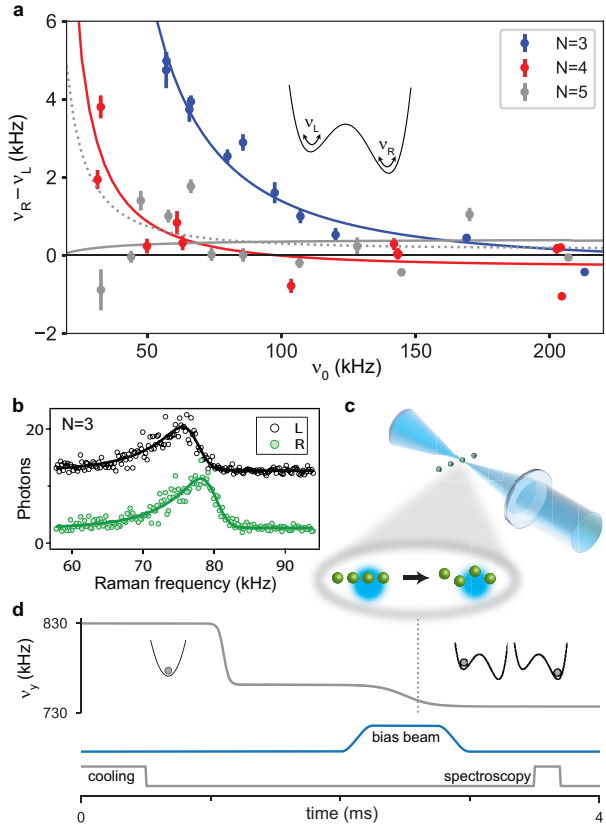}
\caption{\textbf{Spectroscopic characterization of the bias in the linear-zigzag transition.} (a) Difference in zigzag mode frequency, $\protect{\nu_R-\nu_L}$, for an ion crystal of 3--5 ions initialized into opposite zigzag configurations, defined as left (\textit{L}) and right (\textit{R}) sides of the double well. Data plotted as a function of mode frequency $\nu_0$ without any applied initialization. In all cases, $\nu_R\approx\nu_0$ within $0.7$ kHz. Solid lines are fits to classical small-oscillation theory (see text). Bias coefficients from the fits are $C_1=-4.0(3)\times10^{-5}$ and $C_3=-7(7)\times10^{-4}$ for three ions, and $C_1=-9.5(1.7)\times10^{-6}$ and $C_3=1.7(1.4)\times10^{-4}$ for four ions. Data fluctuations for five ions only allow a partial constraint on the biases: a fit of all data (solid gray line) gives $C_1=1(4)\times10^{-5}$ and $C_3=-4(5)\times10^{-3}$, while a fit excluding the left-most data point (dotted line) gives $C_1=-6(5)\times10^{-6}$ and $C_3=-2(3)\times10^{-3}$. The zigzag mode experiences increasing excitation away from the critical point without substantive effect on results expected. (b) Sample lineshapes with fits for three ions at $\nu_0=80$ kHz. (c) Illustration of scheme for initialization into either zigzag structure. Controlled biasing of the transition is achieved with a repulsive optical dipole potential from an off-resonant beam focused with near single-ion resolution. (d) Experiment sequence showing initial laser cooling, ramp of transverse secular frequency $\nu_y$ across the transition biased by the optical potential, and Raman spectroscopy on the zigzag side of the transition to measure $\nu_R$ or $\nu_L$.}
\label{fig:biasSpec}
\end{figure}

To prepare the ion crystal deterministically in either of the zigzag configurations, we use the optical dipole force from a off-resonant laser beam focused to the order of the ion spacing and displaced transversely from a central ion in the string (see Fig.~\ref{fig:biasSpec}c and Methods). This is sufficient to bias the transition globally to achieve initialization of either configuration with $>$90\% fidelity for 3--5 ions. The same experiment sequence as for the transition spectroscopy is used with the addition of the biasing beam applied as the critical point is traversed (Fig.~\ref{fig:biasSpec}(e)). The frequency splitting between the two zigzag configurations is measured as a function of proximity to the critical point on the zigzag side of the transition, and the results are plotted self-consistently in terms of the naturally biased zigzag mode frequency $\nu_0\approx\nu_{R}$ (Fig.~\ref{fig:biasSpec}a). For both three and four ions a rapid reduction of the splitting away from the transition is observed. This is indicative of a dominant linear bias. We fit the measurements to the unapproximated version of Eqn.~\ref{eqn:approx_bias_fit_function_vs_freq} to extract the coefficients $C_1$ and $C_3$ (see Fig.~\ref{fig:biasSpec} for values). The linear bias decreases in magnitude from three to four ions in agreement with the transition spectroscopy of Fig.~\ref{fig:NionDependence}. The value of $C_1$ for three ions, for which the bias is strongest, agrees with that obtained from transition spectroscopy, and the value for four ions agrees within a factor of 2. For five ions, data fluctuations are such that only an upper bound on the small bias coefficients is possible. While this technique is not as precise as a fit to transition spectroscopy, it provides clear evidence of a dominant linear bias for three and four ions, and justifies the fitting of the transition spectroscopy to $C_1$ alone in these cases.

\vspace{0.5em}
\noindent\textbf{Connection of bias coefficients to the trap.} Moving beyond an interpretation of the data in terms of the phenomenological effective potential of Eqn.~\ref{eqn:Ueff}, we have explored theoretically the origin of the zigzag bias in terms of asymmetries in the ion trap potential. The perturbations from an ideal trap can be expressed as a polynomial expansion, $V_{pert}(y_i, z_i) = \sum_{n,m}\lambda_{nm} y_i^{n} z_i^{m}$. A coupled-mode analysis in the pseudopotential limit for perturbations up to 6th order and for ion crystals up to $N=7$ shows that there are only a few trap asymmetries critical to creating the biases $C_1\varphi$ and $C_3\varphi^3$ in the zigzag potential. The linear bias will in general dominate close to the transition where $|\langle \varphi \rangle| \ll 1$. We find that the linear bias is induced by axial-transverse terms $yz^j$ in the trap potential with $j\geq2$ and $j$ even (odd) for an odd (even) number of ions $N$. Importantly, the coefficients $\lambda_{{1,j}}$ are suppressed for $j<N-1$ due to low coupling of the asymmetry to the zigzag mode, which is consistent with the observed rapid suppression of linear bias from 3 to 5 ions. The cubic bias arises from terms $yz^j$ and $y^3z^k$ where $j\geq2$ and $k\geq0$ are even (odd) for odd (even) $N$. Rotations of the trap principal axes and shifts in the trap minimum will induce additional, typically smaller, contributions to the zigzag bias. A linear ion trap potential also has various nonlinearities that do not break zigzag symmetry but can lead to shifts in the critical point, for example quartic terms arising from the endcaps of form $z^4$ and $y^2z^2$ with $\lambda_{{i,j}}$ coefficients in our case of order $10^{-4}-10^{-5}$.

We use numerical modelling of the trap potential as well as estimates of mechanical tolerances to identify likely sources of asymmetry in the trap potential that contribute to the zigzag bias. For our linear trap design (Fig.~\ref{fig:setup}a), the bias for three ions can arise from a number of different trap asymmetries including transverse co-shifts of the endcaps or various displacements of the trap rods~\cite{Chow2022a, Douglas1999a} of order 10 $\mu$m. We find that the same deformations are also roughly consistent with the bias measured for five ions. For four ions, the most likely source of asymmetry is from opposing transverse displacements of the endcaps of order 100 $\mu$m, which is possible given that the endcaps are not positionally well constrained in our trap design.

The theoretical results imply that a highly symmetric zigzag transition can be readily achieved for moderate numbers of ions $N$. With an improved trap with electrodes constrained to a few microns and $N\geq4$, the linear bias $C_1$ can be highly suppressed to below the target $10^{-7}$ level, which we already approach for $N=5$ in our current setup. For odd $N$ this will still leave cubic biases at a level of $<\!10^{-3}$ that decrease only gradually with number of ions. However, ion crystals with even $N$ can be used to advantage to obtain a high degree of double-well symmetry with $|C_3|\lesssim 10^{-6}$, since their transverse symmetry suppresses the effect of the lowest order $\lambda_{{1,2}}$ and $\lambda_{{3,0}}$ terms on the zigzag mode. For example for $N=6$ and in our existing trap, we expect to achieve biases of $C_1=-2.5\times10^{-8}$ and $C_3=1.6\times10^{-5}$.

\section{\label{sec:discussion} Discussion\protect\\}

We have extended previous demonstrations of the LZ transition in linear ion traps to probe the region close to the critical point for ion crystals near the quantum ground state. Stabilized trap potentials, reduced thermal fluctuations and the use of a spectroscopic probe of the zigzag mode have allowed for a precision determination of the critical point beyond previous assessments of the order parameter variation through fluorescence images of the ions~\cite{Enzer2000a, Liang2011a}, and have revealed the effect of small asymmetries in the trap potential on the nature of the transition. We have realized a novel method for measuring the motional population distribution of the zigzag mode near the critical point and demonstrated key ingredients towards the goal of double-well interferometry with this system. A direct extension of this work is to probe the decoherence of the order parameter near the LZ critical point using both a Raman Ramsey measurement of $T_2$ relaxation of the zigzag mode and Raman sideband thermometry to assess the $T_1$ relaxation associated with motional heating. An estimate of the fundamental decoherence limit of image current damping on the zigzag dynamics~\cite{Filippov2011a, Wineland1998a} shows that it is negligible for our ion-electrode distance in comparison to the measured decoherence rate of 0.3--0.5 ms$^{-1}$ near the critical point. Given that we expect the origin of the decoherence is technical in nature, improvements are possible by increasing the ion confinement with smaller trap structures to scale up the characteristic frequencies, and by the reduction of phase noise in the trapping potentials through optimization of the active trap voltage stabilization and the use of additional passive measures, including rf filtering from a higher quality factor rf resonator in a cryogenic trap system.

Extension of this work to a larger number of ions requires a consideration of the scaling of the energy gaps, which impact on diabatic excitation out of the ground state including the formation of spatial domains~\cite{Zurek2005a}. Relevant to this, a calculation of the quantum suppression of long range order in the thermodynamic limit for the case of a homogeneous ion density has been performed~\cite{Shimshoni2011a}. It provides an estimate of a quantum critical point for the LZ transition with a small shift below $\nu_{yc}$, of order 0.1--1 Hz for the central ion density in our trap. A vanishing gap, however, is central to this critical behavior. The non-zero energy gap for a finite number of ions allows for adiabatic evolution of the ground state through the critical point for sufficiently slow quench time~\cite{Zurek2005a}. For our inhomogeneous case in a harmonic trap and out to $\protect{N=50}$, we find theoretically that the location of the optimum tunnel point relative to the critical point $\nu_{yc}$ and the tunnel splitting energy remain nearly constant for a fixed central ion density. As well, the closing of the excitation gap to the next normal mode, which brings with it the possibility of spatial variation in the order parameter, only becomes relevant for very long ion strings, outside the typical 100-ion limit in a linear trap.

Measurements of the decoherence in the zigzag mode provide an assessment of the ambient electric field noise in the trap environment, which is of general relevance to quantum information applications where the ions' motion is used to engineer controlled interactions. While the zigzag mode in a linear ion crystal is less sensitive to fluctuating electric fields than the COM mode due to the higher-order coupling to gradients, the zigzag mode has the useful feature near the LZ transition that it can resolve electric field noise over a wide spectral range from the transverse secular frequency down to dc with minimal variation of the dc trap voltages. In particular this can be performed without change in the rf trap voltage. Furthermore, taking advantage of the high sensitivity of the LZ critical point to both transverse and axial secular confinement, we can gain an enhancement in the measurement of drifts in the trap secular frequencies by using the zigzag mode over a direct measure of the COM modes at the cost of a small additional time overhead required to ramp to near the LZ transition. The single-shot gain over a first-sideband measurement of the transverse COM mode is $G\approx\nu_y/\nu_{zz}$ on the linear side of the transition and reaches a maximum near the critical point due to the energy level structure. For example, we calculate for five ions a gain of up to 67.

Finally, we briefly note additional potential applications of this work. The precise detection of small trap asymmetries with the LZ transition, either via sideband spectroscopy or ultimately with double-well interferometry, should be useful to constrain the effect of trap nonlinearities in quantum computing and quantum simulation with long ion crystals~\cite{Home2011a} and in linear ion trap mass spectrometry~\cite{Douglas1999a}. The spectroscopic sensitivity of the transition and the demonstrated stability of the trap potential open up the possibility to measure finer shifts in the critical point. For example, shifts in the critical point due to cross-mode coupling~\cite{Gong2010a, EjtemaeePhD2015} in the dispersive Kerr regime~\cite{Roos2008a} are small, up to 3 Hz/phonon depending on the contributing mode, but are measurable with the current setup for higher levels of mode excitation. Other potential applications in the quantum regime that build on the techniques presented here include entanglement of the internal spin of the ions and the zigzag mode using a spin-dependent optical dipole force to modify the critical point~\cite{Baltrusch2011a}, and quantum dynamics of topological kink defects seeded into the zigzag structure and prepared in the ground state~\cite{Landa2010a, Timm2021a}.

\clearpage

\section{\label{sec:Methods} Methods\protect\\}

\noindent\textbf{Ion trap system.} The ion trap is a linear radio-frequency (rf) Paul trap~\cite{Ejtemaee2013a,EjtemaeePhD2015} consisting of four rods and two endcap needles with an ion-to-rod distance of $R=0.66$ mm and two needle endcaps with a tip-to-tip separation of 2.5 mm (Fig.~\ref{fig:setup}a). The trap operates at a radio frequency $\Omega_{rf}/2\pi=16.9$ MHz, rf voltage amplitude $V_{rf}=770$ V, and endcap voltage $V_{ec}=67-155$ V depending on ion number. The trap has typical secular frequencies $\protect{\nu_i=\omega_i/2\pi}$ of 864 kHz and 844 kHz along the transverse $x$ and $y$ principal axes and 303 kHz along the axial $z$ axis for five ions starting in the linear phase. Stabilization of the trap potential against drifts is implemented through passive and active techniques that include a servo of the rf amplitude and provide a stability of $<10$ ppm over 100 s for the secular frequencies~\cite{Zhang2022a}. Real-time monitors of the trap rf and endcap voltages provide an assessment of drifts and are used as inputs to a calibrated trap potential model for calculating secular trap frequencies. Excess micromotion is monitored for, but drifts in the background electric field for the majority of datasets are small enough that we can avoid compensation adjustments that could shift the critical point.

We make use of the internal clock states in the $^{171}$Yb$^+$ hyperfine ground state manifold,  $^2S_{1/2}|F=0, m_F=0\rangle\equiv|\downarrow\rangle$ and $\protect{^2S_{1/2}|F=1, m_F=0\rangle\equiv|\uparrow\rangle}$, for ground state cooling and Raman spectroscopy. The general experimental sequence for all data consists of laser cooling the ion string to near the ground state on the linear side well away from the critical point, initialization of the ions into the internal state $|\downarrow\rangle^{\bigotimes N}$, a ramp of the transverse trap frequency to the vicinity of the zigzag transition, application of the Raman sideband spectroscopic probe and finally readout of the internal state of the ions using state selective fluorescence. The total fluorescence from all ions is collected simultaneously onto a single photomultiplier tube.

\vspace{0.5em}
\noindent\textbf{Ground-state laser cooling.} Near ground-state cooling of the ion crystal in the linear-string configuration is achieved through Doppler pre-cooling, 3D Sisyphus cooling of all $3N$ vibrational modes to the few-phonon level~\cite{Ejtemaee2017a}, and finally interleaved resolved sideband cooling of the zigzag mode of interest along the $y$ axis, the $x$-zigzag mode and all other modes along the $y$ and $z$ axes except the center-of-mass (COM) ones. We estimate a ground-state occupation for the sideband cooled modes to be $\gtrsim0.9$ based on measured sideband Rabi oscillations~\cite{Ejtemaee2017a,EjtemaeePhD2015}. Ground-state cooling of modes other than the zigzag mode of interest aids to suppress heating of the zigzag mode due to cross-mode coupling resonances that are encountered during the ramp to the linear-zigzag (LZ) transition.

\vspace{0.5em}
\noindent\textbf{Ramp across the LZ transition.} The approach to and crossing of the LZ transition is controlled by a hyperbolic-tangent ramp of a transverse dc quadrupole potential applied through the trap rods (Fig.~\ref{fig:setup}b)~\cite{Ejtemaee2013a,EjtemaeePhD2015} with minimal effect on the axial confinement ($|\Delta \omega_z|/\omega_z < 0.03\%$). In terms of the quadrupole voltage $V_q$, the secular frequency along the $y$-axis weakens according to $\omega_y = \omega_{y0}\sqrt{1-V_q/V_{q0}}$ for $V_q>0$, and the orthogonal axis strengths as $\omega_x = \omega_{x0}\sqrt{1+V_q/V_{q0}}$, such that the LZ transition is effectively confined to the 2D $y-z$ plane. The quadrupole voltage begins at 0 V and crosses the transition near 2.5 V. For $V_q$ ramp endpoints close to the critical point or those passing into the zigzag phase, a two-stage ramp is used with a slower final stage over 1 ms. Using simulations (see below) we find that this provides sufficient adiabaticity for ramp endpoints near the critical point, but we do not optimize the ramps for endpoints deeper into the zizgag side. An example ramp sequence can be seen in Fig.~\ref{fig:biasSpec}d.

The endpoint of the voltage ramp is converted to a value of transverse secular trap $\nu_y$ using a calibrated trap potential model. We plot results in terms of the control parameter $\nu_y$ rather than the theoretically more natural trap aspect ratio $\alpha$ (Eqn.~\ref{eqn:2Dpotential}) since the axial secular frequency changes only minimally during the quadrupole voltage ramp. The \textit{in-situ} calibration of the trap potential model is derived from prior separate experiments using a single ion and is updated using values of COM modes measured at select points during data collection to correct for daily drifts in the calibration. The trap potential model is a parameterized version of a symmetric linear Paul trap potential up to second order in coordinates and incorporates the next leading order of the Mathieu expansion for the secular frequencies~\cite{CPTvol1}. The calibrations achieve $<0.03$ kHz error from the fits near the critical point. As part of the trap calibration model we implicitly infer the Mathieu $a$ and $q$ parameters~\cite{CPTvol1} for all three trap axes.

\vspace{0.5em}
\noindent\textbf{Raman spectroscopy.} We use motion sensitive two-photon Raman transitions to probe the various vibrational modes in the ion crystal, including the zigzag mode. Different Raman beam pairs allow access to motional sidebands in both the axial and transverse directions. The Raman beams' large size provides nominal uniform illumination of the ion crystal, giving rise to global Raman sideband coupling weighted by the mode participation of each ion~\cite{Wineland1998b}. Since all $N$ ions are simultaneously illuminated by the Raman beams, up to $N$ phonon levels can be excited from a given initial state. While the global coupling complicates the theoretical description of the Raman transition, in practice we keep the average sideband excitation to one phonon or less by limiting the Raman pulse time to simplify the interpretation of measurements. We suppress sideband resonance shifts due to the ac Stark shift of the Raman beams ($\lesssim 0.3$ kHz) by measuring the carrier and sidebands at the same Raman beam power. Near the critical point, reduced Raman beam power helps to resolve close-in sidebands that lie a few kilohertz from the carrier.

\vspace{0.5em}
\noindent\textbf{Bias spectroscopy.} The bias spectroscopy uses the force from an off-resonant beam, focused and offset from a central ion in the linear crystal, to achieve deterministic initialization into either the $L$ or $R$ zigzag configuration before the Raman probe of the zigzag mode is performed. The biasing beam is blue-detuned by 0.7 THz, and its power (1--9 mW) is adjusted to minimize spontaneous emission while still achieving a high fidelity of initialization. The application of the biasing beam is limited to the portion of the quadrupole ramp close to the LZ transition to further minimize spontaneous emission heating. For the data for three and five ions, a controllable temporal amplitude profile has also been implemented to limit non-adiabatic excitation from the bias beam. The fidelity of the zigzag initialization is confirmed over the range of bias spectroscopy data by direct imaging of the zigzag structure following a fast projection ramp deep into the zigzag region. The lineshape asymmetries for the $L$ and $R$ configurations are compared to ensure that the bias beam does not introduce a differential heating effect, which could incur a systematic shift in the resonance line centers. The comparison of zigzag sideband resonances between the $L$ and $R$ configurations is also sensitive to slow drifts in the trap potential. For example, the five-ion bias measurements shown in Fig.~\ref{fig:biasSpec}a are limited by anomalous excess drifts due to electrode contamination.

\vspace{0.5em}
\noindent\textbf{Transition spectroscopy analysis.} To extract the first and second sideband frequencies of the zigzag mode across the LZ transition, as shown in Fig.~\ref{fig:transitionSpec}a and Fig.~\ref{fig:NionDependence}, we fit Raman spectra to a set of incoherently summed peaks without reference to a specific model for the zigzag potential. We assume a Rabi lineshape for the carrier, and for the sidebands we use a Gaussian or exponentially modified Gaussian lineshape depending on proximity to the critical point. The exponentially modified Gaussian is motivated by a model of anharmonicity and a thermal-like distribution of number states for the zigzag mode; however, the functional form is taken as a heuristic function that is found to work well in practice. It is defined as

\begin{widetext}

\begin{equation}
g(f, f_0, w, \Delta_a) =A \sqrt{\pi}\frac{w}{2|\Delta_a|} \exp{\left(-\frac{f-f_0}{\Delta_a}+\left(\frac{w}{2\Delta_a}\right)^2\right)}
\erfc{\left(-\sgn(\Delta_a)\frac{f-f_0}{w}+\frac{w}{2|\Delta_a|}\right)}+b_0
\end{equation}

\end{widetext}

\noindent in terms of frequency $f$, resonance location $f_0$, width $w$, lineshape asymmetry $\Delta_a$, amplitude $A$ and baseline offset $b_0$. The normalization is chosen such that the lineshape area, $Aw\sqrt{\pi}$, matches that of a Gaussian $A\exp(-f^2/w^2)$ with the same width and amplitude.

If the lineshape asymmetry obtained from a fit is below a minimum threshhold ($\Delta_a/w<0.5$) we revert to a Gaussian lineshape. Thus, far from the critical point where anharmoncity is negligible, Gaussians lineshapes are relevant, and they are used again close to the critical point where the sidebands are sufficiently resolved. Exponentially modified Gaussian lineshapes are relevant in the intermediate region where the anharmonicity is significant but not large enough to resolve the sidebands from different initial $n$-number states of the zigzag mode. Identification of the resonances for the first sideband ($n\rightarrow n+1$) and second sideband ($n\rightarrow n+2$) is done by considering continuity of data and by assuming that the initial state is majority populated in the $n\!=\!0$ ground state for data close to the critical point. Close to the critical point on the zigzag side, sideband identification becomes challenging since below $\nu_y-\nu_{yc}=-0.2$ kHz the sidebands for five ions lie under the carrier lineshape or have vanishing line strength, and for four ions the $0\rightarrow1$ and $0\rightarrow2$ resonances approach degeneracy. In the case of five ions below $-0.2$ kHz, low-frequency peaks extracted from underneath the carrier lineshape have been included conservatively as the upper first sideband $0 \rightarrow 1$ in Fig.~\ref{fig:transitionSpec}a but may be due to the lower first sideband $1\rightarrow0$ or an upper sideband $n\rightarrow n+1$ for $n>0$.

\vspace{0.5em}
\noindent\textbf{LZ transition fit.} We fit the sideband frequency dependence across the LZ transition in an iterative fashion to obtain the critical point $\nu_{yc}$ and the bias in the zigzag potential. The fitting procedure is as follows: (i) An initial value of $\nu_{yc}$ is extracted from a fit to the ideal classical dependence of the zigzag mode on the linear side, $\nu_{zz}=\sqrt{\nu_y^2-\nu_{yc}^2}$, for a data range limited to $\gtrsim$1 kHz above the critical point. This avoids bias and quantum effects that manifest near the critical point. (ii) An initial value for the bias is determined from a fit to the numerical quantum theory for the zigzag potential (Eqn.~\ref{eqn:Ueff}) using $\nu_{yc}$ from (i). (iii) A refined value of the critical point is determined from a fit of each sideband to the quantum theory on the linear side of the transition and the results are averaged. (iv) A refined value for the bias is determined from a fit of both sidebands to the quantum model. Refinement of the critical point is less than 0.05 kHz for the data presented. Statistical error in the critical point from fitting and systematic error from determination of the secular trap frequencies are $<0.02$ and $<0.03$ kHz respectively for $\nu_{yc}=717-760$ kHz. Critical trap asymmetries $\alpha_c$ are calculated using the measured axial secular frequency. The data does not allow to distinguish a linear from cubic bias. Using additional information from bias spectroscopy, we determine for three and four ions that a linear bias dominates, and so fits to a linear bias with $C_3=0$ are presented in Fig.~\ref{fig:NionDependence}. For five ions the bias is small enough that the sideband frequency dependence alone can only roughly constrain the bias. An alternative fitting approach is taken to constrain the bias further (see discussion below).

For higher data density in spectroscopy plots of the LZ transition (Figs.~\ref{fig:transitionSpec}a and~\ref{fig:NionDependence} for four and five ions), we take advantage of the stability of the experiment to combine data sets from two or more days. Data sets are combined by fitting each set for the critical point using the above method and then using the fit value as the reference for the horizontal axis, \protect{$\nu_y-\nu_{yc}$}, to remove small day-to-day drifts. The critical points quoted in the main text represent a weighted statistical average over values from all data sets.

\vspace{0.5em}
\noindent\textbf{Motional state and bias assessment from line centers and strengths.} For the five-ion Raman spectra close to the critical point where the sideband peaks from different initial number states of the zigzag mode are resolved or partially resolved, we use the relative peak heights and center frequencies of the sidebands to extract both the motional population distribution of the number states and the bias in the zigzag potential. We fit spectra including the carrier, multiple upper sidebands and the first lower sideband as an incoherent sum of lineshapes. Refined models of the lineshapes are incorporated into the fit: both the carrier and sidebands are modelled as Rabi lineshapes with pure phase damping. The lineshapes are obtained from numerical solution of the optical Bloch equations~\cite{Metcalf}, and in practice lie between the limits of Rabi and Lorentzian forms. The value of the sideband Rabi couplings are calculated using (i) wavefunctions obtained for a given zigzag potential, (ii) a global Raman interaction arising from uniform laser illumination of the ions, and (iii) \textit{no} assumption of the Lamb-Dicke approximation~\cite{Wineland1998b}. For a resolved sideband from an initial $n$-number state, the Raman transition with global coupling simplifies to that for a two-level system, $|\downarrow\rangle^{\bigotimes \!N}|n\rangle \leftrightarrow |\Uparrow_{\mathit{eff}}\rangle|m\rangle$ where $|\Uparrow_{\mathit{eff}}\rangle$ is a superposition of single-excitation spin states. The effective Rabi frequency is

\begin{equation}
\Omega_{nm} = \Omega_0 \sqrt{\sum_{j=1}^{N}{|M_{nm,j}|^2}}
\end{equation}

\noindent where the matrix element $M_{nm,j}$ for the $j^{th}$ ion is

\begin{equation}
M_{nm,j} = \langle \Psi_n(\varphi) |e^{(i k_{\mathit{eff}\!,j}a_z\varphi)}|\Psi_m(\varphi) \rangle
\end{equation}

\noindent in terms of zigzag wavefunctions $|\Psi_n\rangle$, scaled zigzag normal mode coordinate $\varphi$, Coulomb length scale $a_z$, and effective wavenumber
$\protect{k_{\mathit{eff},j}= \vec{k}_{L} \cdot \hat{D}_{\mathit{zz}}^{j}}$, which
captures the projection of the Raman wavevector $\vec{k}_L$ onto the direction $\hat{D}_{\mathit{zz}}^{j}$ of the $j^{th}$ ion's displacement in the zigzag mode. The wavefunctions are obtained from numerical solution of the time-independent Schrodinger equation for the zigzag potential (see below). An approximate linear model for the axial components of the zigzag mode vector is used on the zigzag side of the transition. The model works well out to a zigzag mode frequency of 60 kHz on the zigzag side, which is adequate for the data range of interest.

Fit parameters for a measured carrier and sideband spectrum include a baseline offset, offset from the critical point $\protect{\nu_y-\nu_{yc}}$, potential parameters $C_1$ and $C_3$, a carrier Rabi frequency $\Omega_0$ that sets the overall coupling scale, a carrier phase damping rate $\gamma_{car}$ and carrier amplitude correction factor near unity, first and second sideband phase damping rates $\gamma_1$ and $\gamma_2$ and an overall relative amplitude correction to the carrier, and finally motional populations $P(n)$ for the number states of the zigzag mode. Fit parameters and uncertainties are obtained from unweighted fits using $\chi^2$ minimization.

We find that $C_1$ and $C_3$ values from the fits are strongly correlated, but the motional populations $P(n)$ are insensitive to the choice. Without additional constraining information, we choose $C_3=0$ and determine an average value of $C_1=3.3(3)\times10^{-7}$ from the resulting fits  to the data. We use this value to obtain final fit values for the motional distributions $P(n)$ as shown in Fig.~\ref{fig:transitionSpec}. The fits could be improved by including a set of lower sideband measurements but at the expense of longer data collection times.

\vspace{0.5em}
\noindent\textbf{Classical small-oscillation analysis.} Small-oscillation analysis is performed for an $N$-ion crystal in the 2D pseudo-potential (Eqn.~\ref{eqn:2Dpotential}) to extract the equilibrium crystal structure and normal modes. The critical value $\alpha_c^{(0)}$ for the LZ transition in the pseudopotential approximation is extracted from the eigenvalue spectrum of the normal modes~\cite{Enzer2000a}. Values for 3--5 ions in an ideal linear trap are $\protect{\alpha_c^{(0)}=\{2.4000, 4.1542, 6.2374\}}$. The frequency of the zigzag mode experiences a shift from the pseudo-potential value due to micromotion effects in the ion trap~\cite{Landa2012a, Kaufmann2012b}, which in turn leads to a shift in the critical point at the percent level. We apply approximate Mathieu equations for the normal modes of a linear ion string~\cite{Landa2012a} to derive the micromotion-corrected frequency for the zigzag mode and the micromotion-shifted critical point

\begin{equation}
\alpha_c\approx\alpha_c^{(0)}\left[1+\frac{q_y^2}{2}
\left(1+\frac{3}{8}q_y^2+\frac{5\omega_{\mathit{z}}^2}{2\Omega_{\mathit{rf}}^2}\alpha_c^{(0)}\right)\right]
\end{equation}

\noindent in terms of the Mathieu parameter $q_y$ along the relevant transverse $y$-axis. For this result we assume $q_y^2\ll1$, neglect small corrections due to Mathieu parameter $a_y$, and neglect weak $\omega_y$ dependence by setting $\omega_y^2\approx\omega_z^2\alpha_c^{(0)}$. Since the lowest order effect on the zigzag mode for linear or nearly linear ion crystals is a frequency shift, we use the pseudopotential result for the small-oscillation curve shown in Fig.~\ref{fig:transitionSpec}a. Far from the critical point on the zigzag side of the transition, the micromotion-corrected frequency of the zigzag mode needs to be determined numerically from the crystal structure and normal mode vector~\cite{Landa2012a, Kaufmann2012b}, but this is not necessary for this work.

\vspace{0.5em}
\noindent\textbf{Zigzag potential.} To derive the effective potential for the zigzag mode near the critical point we perform a classical, perturbative, coupled-mode analysis and adiabatically eliminate modes other than the soft zigzag mode. Using this semi-analytical method, we can obtain the sensitivity of the potential coefficients $C_1$ through $C_4$ (Eqn.~\ref{eqn:Ueff}) to trap perturbations $\lambda_{i,j}y^iz^j$ in $V_{pert}$ (Eqn.~\ref{eqn:2Dpotential}). We briefly summarize the method with further details to be presented elsewhere~\cite{Chow2022a}. We Taylor-expand the pseudo-potential (Eqn.~\ref{eqn:2Dpotential}) to order $\geq5$ in the ions' displacements and express the result in terms of the mode coordinates $\varphi_i$ of the linear crystal's $2N$ normal modes, which are re-diagonalized at lowest order of coupling to account for the trap perturbations $V_{pert}$. Working from the nonlinear coupled oscillator equations near the critical point, we assume that all other modes adiabatically follow the soft zigzag mode and so adiabatically eliminate them to arrive at an effective force for the zigzag mode in terms of the zigzag order parameter, $\varphi\equiv\varphi_{zz}$, alone. Integration of the force provides the effective zigzag potential, which is limited to $4^{th}$ order. We retain up to $1^{st}$ order in the polynomial coefficients of $V_{pert}$, which give rise to bias coefficients $C_1$ and $C_3$, shifts in the critical point and modifications to $C_4$. The value of $C_4$ also includes the effect of axial collapse of the ion crystal as the zigzag structure grows, which is encoded in the coupling between the zigzag and axial breathing modes. The effect is significant (order unity) and its inclusion extends the double-well calculation performed in~\cite{Retzker2008a}. Values of $C_4$ in an ideal linear trap are $\{0.9307,2.1906,4.5409\}$ for 3--5 ions. We ignore in our fits to data the small effects of the trap perturbations on $C_2$ (i.e. on the critical point) and on $C_4$. Also in our calculation the coupling to the zigzag mode is limited to the effect of the mean order parameter for the non-zigzag modes, $\langle\varphi_{i\neq zz}\rangle$, and excludes the effect of fluctuations arising from terms $\langle\varphi_{i}^2\rangle$ -- for example due to thermal fluctuations -- that induce a shift in the critical point~\cite{EjtemaeePhD2015}. We expect these effects to be small given the near-ground-state initial cooling of the relevant modes.

\vspace{0.5em}
\noindent\textbf{Quantum theory.} We assume the quartic form of the effective zigzag potential $U(\varphi)$ as defined in Eqn.~\ref{eqn:Ueff} as the starting point for quantum mechanical calculations of the zigzag mode. This implicitly ignores axial spatial variation of the zigzag order parameter $\varphi$~\cite{Shimshoni2011a}, which applies for the small ion crystals being considered. The effective zigzag potential, derived from the classical coupled mode theory, also ignores the effect of quantum fluctuations from the other modes, which is reasonable given their high frequency. The values of $\alpha_c$ and $C_4$ in the potential are obtained from the classical theory in the absence of trap imperfections, and the bias coefficients $C_1$ and $C_3$ are treated as adjustable model parameters obtained from fits. The quantized energy levels and zigzag wavefunctions $\Psi(\varphi)$, used in the preparation of Figs.~\ref{fig:setup}--\ref{fig:NionDependence}, are obtained as a function of $\alpha$ from a numerical solution of the 1D time-independent Schrodinger equation for the potential $U(\varphi)$. The wavefunctions are used to calculate Raman sideband couplings, as discussed above. Adiabaticity of the ramp across the transition (Fig.~\ref{fig:transitionSpec}h) is assessed from numerical solution of the 1D time-dependent Schrodinger equation in which the applied quadrupole voltage ramp introduces a time varying trap aspect ratio $\alpha(t)$ to the zigzag potential. The form of the quartic zigzag potential derived perturbatively near the critical point begins to break down significantly for $\gtrsim100$-kHz mode frequency on the zigzag side, which provides sufficient range for comparison to experiment.

\vspace{0.5em}
\noindent\textbf{Trap potential simulation.} To understand the role of the trap electrode imperfections in generating zigzag bias, we simulate the electric potentials from the trap electrodes using commercial Charged Particle Optics (CPO) software~\cite{cpo}. For a given trap electrode configuration, the potentials are least-squares fit to a set of spherical harmonics, which are subsequently used to obtain the 3-D Cartesian coefficients $\lambda_{i,j,k}$ for the $x^iy^jz^k$ polynomial expansion of the trap potential in the pseudopotential approximation. The Cartesian coefficients are expressed in a coordinate basis with origin defined by ideal micromotion compensation and with axes defined by the principal axes of the pseudopotential. In conjunction with the sensitivities of the zigzag biases $C_1$ and $C_3$ to the $\lambda$ coefficients (see above), we assess the role and magnitude of various electrode deformations in generating bias in the zigzag potential~\cite{Chow2022a}. Assuming that the lowest-order asymmetry is the dominant contribution, the linear bias coefficient for three ions is $C_1\approx-0.95\,\lambda_{12}$. This can arise from a number of different rod and endcap deformations of order 10 $\mu$m. The same deformations make contributions to $C_3$ of order $10^{-3}-10^{-4}$. They are also reasonably consistent with the linear bias measured for five ions, where $C_1 \approx -7.6\cdot10^{-4}\lambda_{{1,2}}+1.50\lambda_{{1,4}}$. For four ions, the linear bias at lowest order of perturbation is $C_{1} \approx 1.14\,\lambda_{13}$, which is likely due to transverse shifts in the endcap positions of the trap.

\section{\label{sec:acknowledgements} Acknowledgements\protect\\}
\begin{acknowledgments}
We thank Haggai Landa for helpful discussions. This work was supported by NSERC and CFI LOF.
\end{acknowledgments}

\section{\label{sec:author_contributions} Author contributions\protect\\}
SE and PCH designed the initial experiment. All authors contributed to setup, data collection and analysis over the course of the project, and all authors contributed to the preparation of the final manuscript.

\section{\label{sec:competing_interests} Competing interests\protect\\}
The authors declare no competing interests.


%




\end{document}